\documentclass[aps,preprint,showpacs]{revtex4}

\usepackage{amssymb}
\usepackage{amsmath}
\usepackage{epsfig}
\usepackage{epstopdf}
\usepackage{bm}
\usepackage{graphicx,epsfig}
\usepackage{mathrsfs}
\usepackage{dcolumn}
\usepackage{color}
\usepackage{natbib}
\usepackage{CJK}
\def\be{\begin{equation}}
\def\ee{\end{equation}}
\def\bea{\begin{eqnarray}}
\def\eea{\end{eqnarray}}

\allowdisplaybreaks[2]

\begin{document}

\title{Topologically non-trivial Floquet band structure in a system undergoing photonic transitions in the ultra-strong coupling regime}
\author{Luqi Yuan and Shanhui Fan}
\affiliation{Department of Electrical Engineering, and Ginzton
Laboratory, Stanford University, Stanford, CA 94305, USA}

\date{\today }

\begin{abstract}
We consider a system of dynamically-modulated photonic resonator
lattice undergoing photonic transition, and show that in the
ultra-strong coupling regime such a lattice can exhibit
non-trivial topological properties, including topologically
non-trivial band gaps, and the associated topologically-robust
one-way edge states.  Compared with the same system operating in
the regime where the rotating wave approximation is valid,
operating the system in the ultra-strong coupling regime results
in one-way edge modes that has a larger bandwidth, and is less
susceptible to loss. Also, in the ultra-strong coupling regime,
the system undergoes a topological insulator-to-metal phase
transition as one varies the modulation strength. This phase
transition has no counter part in systems satisfying the rotating
wave approximation, and its nature is directly related to the
non-trivial topology of the quasi-energy space.
\end{abstract}


\maketitle

Creating topological effects \cite{klitzing80} for both electrons
\cite{qi11} and photons
\cite{lu14,haldane08,raghu08,wang09,hafezi11,poo11,hafezi13,khanikaev13,longhi13,liang13,mittal14,li14,slobozhanyuk15}
are of significant current interests. A powerful mechanism for
achieving non-trivial topology in a system is to dynamically
modulate the system in time. For electrons, such modulation can be
achieved by coupling with an external electromagnetic field, and
has been used to create an electronic Floquet topological
insulator
\cite{kitagawa10,inoue10,lindner11,leon13,rudner13,cayssol13}. For
photons, optical analogue of Floquet topological insulators has
been demonstrated \cite{rechstman13}. Also, time-dependent
refractive-index modulations can be used to create an effective
magnetic field \cite{fang12np}, which can break time-reversal
symmetry and create a one-way edge state that is topologically
protected against arbitrary disorders.

All previous works on the topological behaviors of dynamically
modulated systems have considered only the \textit{weak-coupling}
regime where the modulation strength is far less than the
modulation frequency, and applied the rotating wave approximation
(RWA). On the other hand, in recent years, the study of
light-matter interactions in the \textit{ultra-strong coupling}
regime, where the rotating wave approximation is no longer valid,
is becoming important
\cite{irish05,irish07,rabl11,schiro12,burillo14,gunter09,imamoglu09,niemczyk10}.
It is therefore of fundamental importance to understand
topological effects beyond the rotating wave approximation.

In this Letter, we analyze a time-dependent Hamiltonian first
proposed in Ref. \cite{fang12np} for achieving an effective
magnetic field for photons. Unlike Ref. \cite{fang12np}, however,
here we focus the ultra-strong coupling regime where the rotating
wave approximation is no longer valid. Experimentally, reaching
such ultra-strong coupling regime is in fact relatively
straightforward with current photonic technology \cite{tzuang14}.
For this system in the ultra-strong coupling regime, we show that
the topologically protected one-way photonic edge states can
persist over a broad parameter range. Compared with the
weak-coupling regime, the topologically protected one-way edge
state is less susceptible to intrinsic losses. We also show that,
as one varies the modulation strength, there is a topological
phase transition that is uniquely associated with the ultra-strong
coupling regime, and has no counter part in weak-coupling systems.

\begin{figure}[h]
\includegraphics[width=10cm]{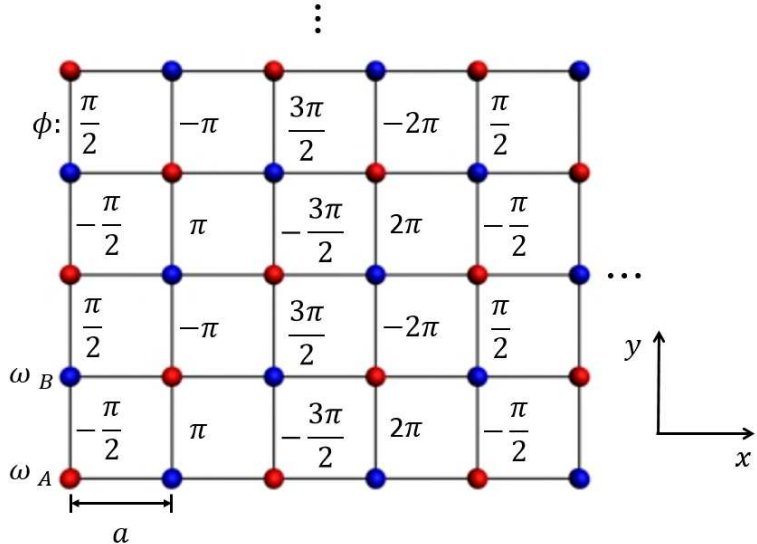}
\caption{Lattice composed of two types of resonators $A$ (red) and
$B$ (blue). The lines represent coupling (or bonds) between
nearest neighbors. All coupling strengths are modulated in time
harmonically. For all bonds along the horizontal direction the
modulation phase is zero. The bonds along the vertical direction
have a spatial distribution of modulation phases as specified in
the figure. The specified phases are for the hopping matrix
elements of the Hamiltonian in Eq. (\ref{ham1}) along the positive
$y$ direction.} \label{lattice}
\end{figure}

We start with the Hamiltonian \cite{fang12np}
\begin{equation}
H = \omega_A \sum_m a^\dagger_m a_m + \omega_B \sum_n b^\dagger_n
b_n  + \sum_{\langle mn \rangle}V\cos(\Omega t + \phi_{mn})
(a^\dagger_m b_n+b^\dagger_n a_m), \label{ham1}
\end{equation}
which describes a lattice of photonic resonators as shown in
Figure \ref{lattice}. The lattice consists of two sub-lattices,
each consisting of resonators of resonant frequencies $\omega_A$
and $\omega_B$, respectively. $a^\dagger$ ($a$) and $b^\dagger$
($b$) are the creation (annihilation) operators associated with
the resonators in the two sub-lattices. The coupling between the
resonators are modulated dynamically, where $V$ is the maximum
coupling strength, $\Omega = \omega_A - \omega_B$ is the
modulation frequency, and $\phi_{mn}$ is the modulation phase.
Such a modulation drives a photonic transition \cite{winn99}
between nearest neighbor resonators.

Eq. (\ref{ham1}) can be transformed to the rotating frame with
\begin{equation}
\tilde a_m(\tilde b_n) = a_m(b_n)e^{i\omega_{A(B)}t}.
\label{gauge}
\end{equation}
The Hamiltonian then becomes
\begin{equation}
\tilde H = \frac{V}{2} \sum_{\langle mn \rangle} (\tilde
a^\dagger_m \tilde b_n e^{-i\phi_{mn}}+\tilde b^\dagger_n \tilde
a_m e^{i\phi_{mn}} + \tilde a^\dagger_m \tilde b_n e^{i2\Omega
t+i\phi_{mn}}+\tilde b^\dagger_n \tilde a_m e^{-i2\Omega
t-i\phi_{mn}}), \label{ham2}
\end{equation}
where the first two terms define the Hamiltonian $\tilde
H_{\mathrm{RWA}}$ in the rotating wave approximation, and the last
two terms are commonly referred as the counter-rotating terms.

Ref. \cite{fang12np} considered the weak-coupling regime where
$V\ll\Omega$ and applied the rotating wave approximation by
ignoring the counter-rotating terms in Eq. (\ref{ham2}). In this
case, $\tilde H \simeq \tilde H_{\mathrm{RWA}}$ becomes
time-independent, and is identical to the Hamiltonian of a quantum
particle on a lattice subject to a magnetic field described by a
vector potential $\bm{A}$, with $\phi_{mn} = \int_m^n \bm{A} \cdot
d \bm{l}$ \cite{peierls33}. Such an effective magnetic field
provides a new mechanism for controling the propagation of light
\cite{fang13,fang13oe,lin14,yuan15}. In particular, it can be used
to achieve a topologically protected one-way edge state
\cite{fang12np}.

However, from an experimental point of view, it is important to
explore the topological behavior of the Hamiltonian of Eq.
(\ref{ham1}) beyond the weak-coupling regime. The required
modulation in Eq. (\ref{ham1}) can be achieved electro-optically
\cite{tzuang14}, in which case $V \sim \delta n/n \cdot \omega_0$,
where $n$ is the refractive index, $\delta n$ is the strength of
the index modulation, and $\omega_0$ is the operating frequency.
For electro-optic modulation of silicon, to minimize free-carrier
loss, $\delta n/n \approx 10^{-5} - 10^{-4}$ \cite{lipson05}.
Optical communication typically uses an $\omega_0$ corresponding
to a free-space wavelength near 1.5 micron. Thus $V$ is typically
between 10-100 GHz. On the other hand, the modulation frequency
$\Omega$ in electro-optic modulation is also on the order of
10-100 GHz \cite{tzuang14}. Thus, experimentally one can readily
operate in the regime with $V \sim \Omega$. Similar conclusion can
be reached for the acoustic-optical modulation scheme implemented
in Ref. \cite{li14} for achieving a photonic gauge potential.
Unlike the electronic transition, where reaching the ultra-strong
coupling regime is a significant challenge
\cite{scalari12,geiser12,li13,scalari14}, for photonic transition
\cite{winn99} it is in fact rather natural that the system
operates in the ultra-strong coupling regime. Thus, systems
exhibiting photonic transition can be readily used to explore the
physics of ultra-strong coupling.

To explore the topological properties of Eq. (\ref{ham1}) in the
ultra-strong coupling regime, we perform a Floquet analysis of the
Hamiltonian of Eq. (\ref{ham2}). Here we choose a spatial
distribution of the modulation phase as shown in Figure
\ref{lattice}. All bonds along the horizontal direction have a
zero modulation phase. The bonds along the vertical direction have
a spatial distribution of modulation phases. In the weak-coupling
regime, such a distribution corresponds to an effective magnetic
flux of $\pi/2$, or 1/4 of the magnetic flux quanta per unit cell.
The system is therefore topologically non-trivial in the
weak-coupling regime. The aim of the Floquet analysis is then to
see to what extent such non-trivial topological feature persist as
one goes beyond the rotating wave approximation.

Our Floquet band structure analysis follows that of Refs.
\cite{shirley65,samba73}. The system in Figure \ref{lattice} is
periodic spatially. Therefore the Hamiltonian can be written in
the wavevector space ($\bm{k}$-space). For each $\bm{k}$ point,
$\tilde H$ in Eq. (\ref{ham2}) has a period in time
$T=\pi/\Omega$. Therefore, the solution of the equation $i d/dt
|\Psi\rangle - \tilde H |\Psi\rangle = 0$ in general takes the
form $|\Psi\rangle = e^{-i\varepsilon t}|\Phi\rangle$, where
$|\Phi (t+T) \rangle = |\Phi (t) \rangle$ has a periodicity $T$ in
time, and is commonly referred to as the Floquet eigenstate.
$\varepsilon$ is the quasi-energy and is defined in the temporal
first Brillouin zone $\varepsilon \in [-\pi/T, \pi/T] = [-\Omega,
\Omega]$. The quasi-energies and the Floquet eigenstates can be
obtained by solving numerically the eigenvalue equation: $(\tilde
H - i\partial /\partial t) |\Phi\rangle = \varepsilon
|\Phi\rangle$ \cite{shirley65,samba73}. The quasi energy thus
obtained as a function of $\bm{k}$ defines the Floquet band
structure.

As an important subtlety, when performing numerical calculation of
the Floquet band structures, it is essential to perform a gauge
transformation such that the resulting Hamiltonian has the
smallest possible temporal period. For our case here, the
transformation of Eq. (\ref{gauge}), which is a gauge
transformation, serves this purpose. While the Hamiltonians $H$
(in Eq. (\ref{ham1})) and $\tilde H$ (in Eq. (\ref{ham2})) are
equivalent to each other since they are related by the gauge
transformation of Eq. (\ref{gauge}), the temporal periods of $H$
and $\tilde H$ are $2\pi/\Omega$ and $\pi/\Omega$, respectively. A
key aspect of topological band structure analysis is to identify
band gaps that are topologically non-trivial. On the other hand,
if one analyze $H$ directly, since the corresponding temporal
first Brillouin zone is smaller, there is additional band folding
along the quasi-energy axis, which obscures the band gap. The use
of $\tilde H$ in Eq. (\ref{ham2}) is in fact quite important for
the analysis of the topological aspects of the Floquet band
structure.

\begin{figure}[h]
\includegraphics[width=14cm]{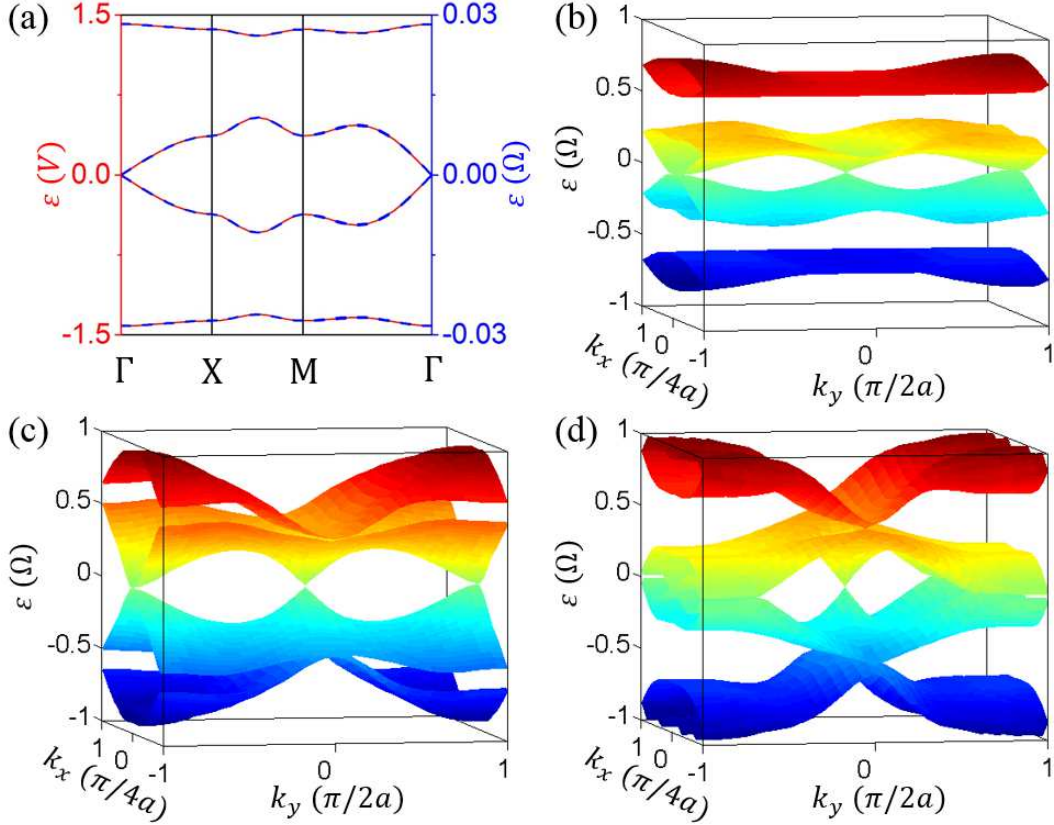}
\caption{Floquet band structure for the Hamiltonian $\tilde H$ of
Eq. (\ref{ham2}), (a) with RWA from $\tilde H_{\mathrm{RWA}}$ (red
solid line) and $V=0.02\Omega$ (blue dashed line), (b)
$V=0.5\Omega$, (c) $V=1.1\Omega$, and (d) $V=1.5\Omega$.
\label{fbsrot}}
\end{figure}

We now examine the Floquet band structure of the system. $\tilde
H$ has a spatial periodicity of 4$a$ by 2$a$ along the $x$ and $y$
directions, respectively. Thus, its Floquet band structure has 8
bands, as we can see in Figure \ref{fbsrot}.

As a comparison, we first consider the band structure of the
Hamiltonian $\tilde H_{\mathrm{RWA}}$ as defined by ignoring the
counter-rotating terms in Eq. (\ref{ham2}).  $\tilde
H_{\mathrm{RWA}}$ has a spatial periodicity of 4$a$ by 1$a$ along
the $x$ and $y$ directions, respectively. However, to facilitate
the comparison with the band structure of $\tilde H$, here we plot
the bandstructure of $\tilde H_{\mathrm{RWA}}$ with the spatial
periodicity of 4$a$ by 2$a$ as well. The resulting RWA band
structure, as shown in Figure \ref{fbsrot}a, thus contains 8
bands. The bands are two fold degenerate. The bandstructure has
the same shape for different values of $V$. (Figure
\ref{fbsrot}a).

In the RWA band structure, there are two gaps separating the
middle group of four bands from the upper and the lower groups,
each of two bands, respectively. These gaps are topologically
non-trivial, as can be checked by calculating the Chern number
\cite{rudner13}:
\begin{equation}
\mathcal{C} = -\frac{1}{2\pi} \sum_\alpha \int dk_x dk_y
(\nabla_{\vec{k}} \times \mathcal{A}_{\alpha}), \label{chern}
\end{equation}
where the summation is over the group of bands, each band indiced
by a different $\alpha$.
\begin{equation}
\mathcal{A}_{\alpha} = \langle \Phi_\alpha
(\vec{k},t)|i\nabla_{\vec{k}} |\Phi_\alpha (\vec{k},t) \rangle,
\end{equation}
The Chern numbers for the upper, middle and lower groups of bands
are +1, -2, +1, respectively. (As a side note, the middle group of
four bands can actually be separated into two subgroups each
consisting of two bands, separated by Dirac points at
$\varepsilon=0$. Each of the subgroup has a Chern number -1.) The
topological analysis here is consistent with the association of an
effective magnetic field in this system. The gaps remain open for
all non-zero values of $V$. Thus with the rotating wave
approximation there is no phase transition as one varies $V$.

Having reviewed the band structure of $\tilde H_{\mathrm{RWA}}$,
we now consider the Floquet band structure of the full Hamiltonian
$\tilde H$. Figure \ref{fbsrot}a shows the cases of $V = 0.02
\Omega$, the Floquet band structure $\tilde H$ agrees very well
with that of the $\tilde H_{\mathrm{RWA}}$.

As one increases $V$ to approximately $V > 0.1 \Omega$, the RWA is
no longer adequate to describe the band structure of $\tilde H$.
Hence, it is no longer possible to interpret the bandstructure
using the concept of an effective magnetic field. Nevertheless,
the two gaps remain open for $V$ ranging from near zero to
$1.1\Omega$ (see Figures \ref{fbsrot}). Therefore, in this range
of $V$, the Chern numbers for the upper, middle and lower groups
of bands must remain unchanged at +1, -2, +1, respectively, and
hence the topological aspects of the band structure remains the
same as one goes into the ultra-strong coupling regime.

As $V$ increases from 0, the bands gradually move away from
$\varepsilon = 0$, and start to occupies more of the temporal
first Brillouin zone (See Figure \ref{fbsrot}b). At $V \sim
\Omega$, some of these bands reach the edge of the temporal first
Brillouin zone. Further increase of $V$ then results in the
folding back of these bands back into the temporal first Brillouin
zone and the closing of the gaps as we see in Figure \ref{fbsrot}c
with $V=1.1\Omega$. No gap is found for larger values of $V$. We
see that the increase of $V$ induces a topological phase
transition: the system behaves as a Floquet topological insulator
at small $V$, and a gapless and topologically trivial ``metal'' at
large $V$.

\begin{figure}[h]
\includegraphics[width=8cm]{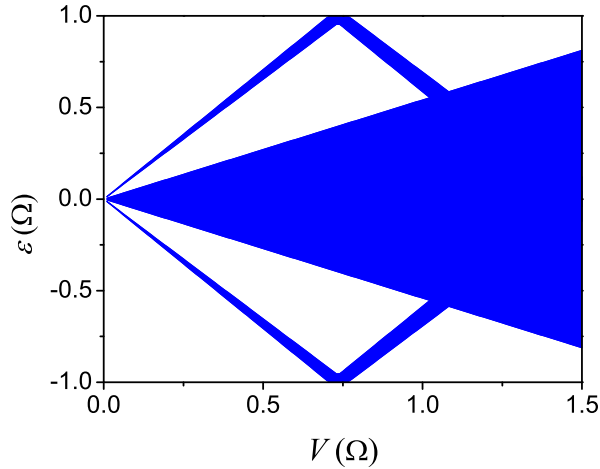}
\caption{The projected RWA band structure folded into the temporal
first Brillouin zone as a function of $V$. The blue regions
correspond to the bands. The white regions are the band gap
regions. \label{fbsrwa}}
\end{figure}

The values of $V$ required for achieving such a topological phase
transition can be estimated by folding the RWA band structure into
the temporal first Brillouin zone of [$-\Omega$, $\Omega$] (Figure
\ref{fbsrwa}), which predicts that the gap closes at $V =
1.1\Omega$. In comparsion, for the full Hamilontian of Eq.
(\ref{ham2}), the gap actually closes at $V = 1.11 \Omega$.
Therefore, We see that this topological phase transition is
directly related to the non-trivial topology of the quasi-energy
space.

\begin{figure}[h]
\includegraphics[width=8cm]{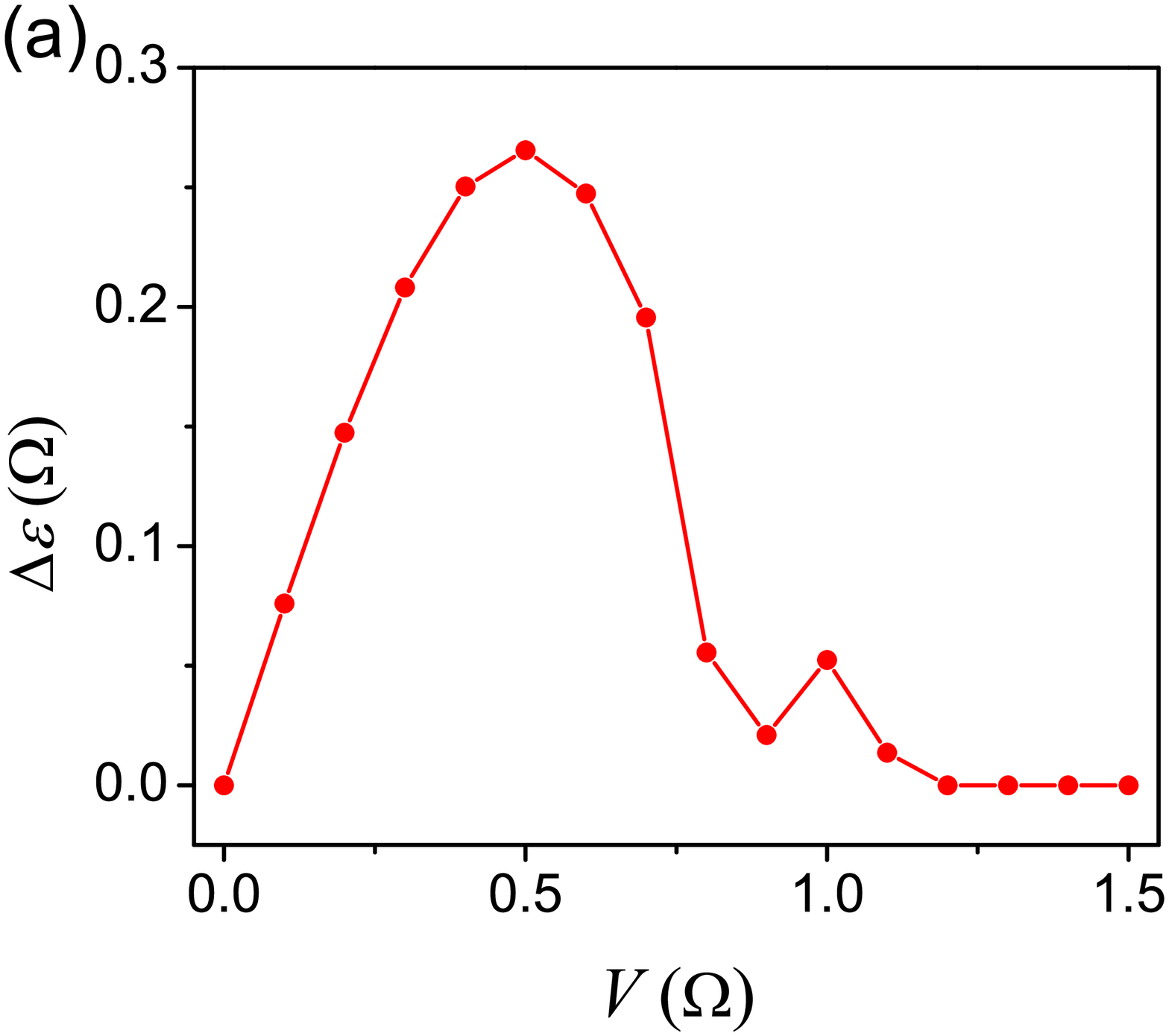}
\includegraphics[width=8cm]{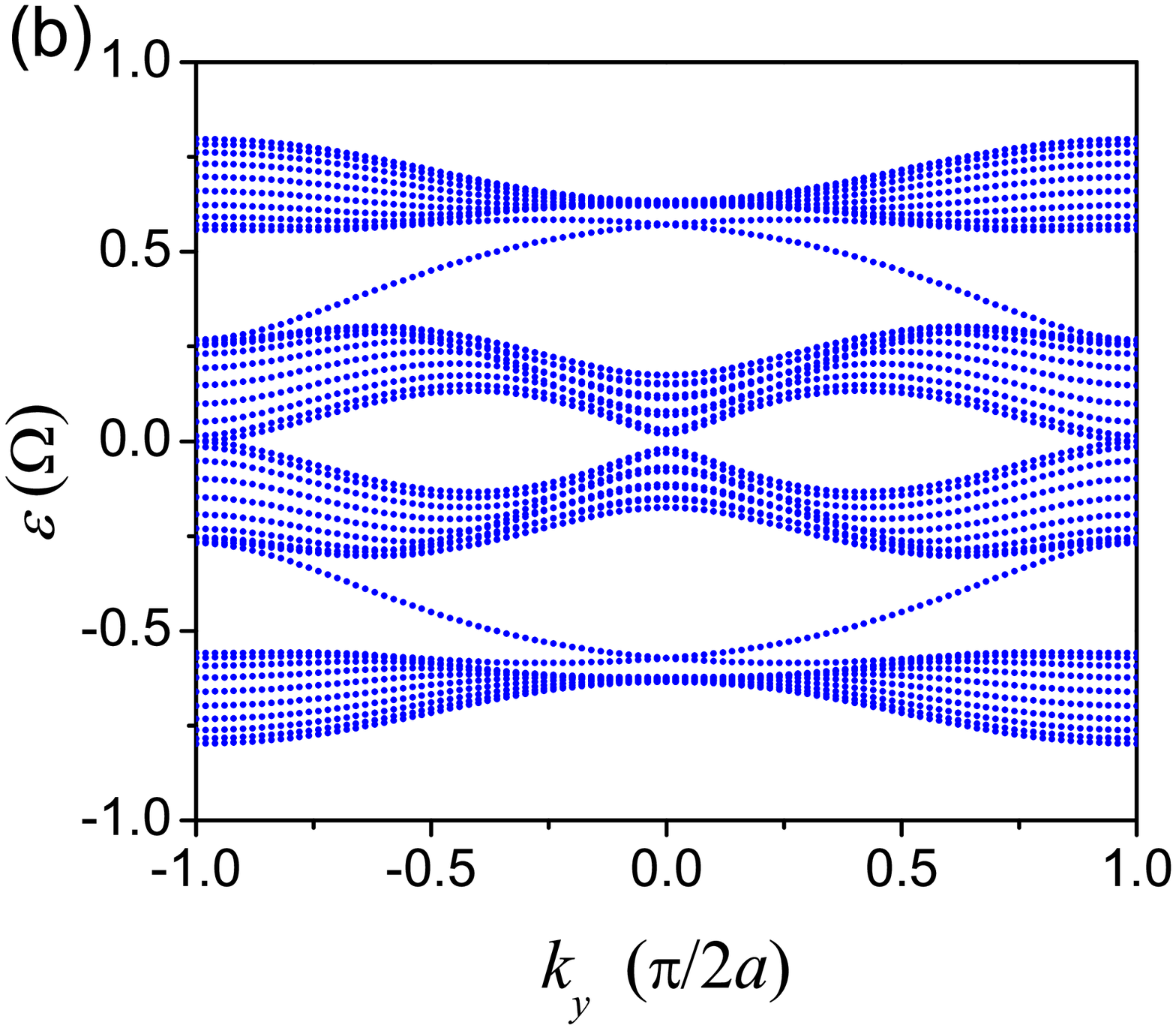}
\caption{(a) The bandwidth of the topologically non-trivial gap as
a function of $V$. (b) The projected bandstructure at
$V=0.5\Omega$. \label{edge}}
\end{figure}

A key signature of a topologically non-trivial band gap is the
existence of a one-way edge state in a strip geometry. We
therefore calculate the projected Floquet band structure for a
strip that is infinite in the $y$-direction and has a width of
$21a$ in the $x$-direction. The projected band structure consists
of the quasi energy of all the eigenstates of the system as a
function of $k_y$. The projected band consists of three groups of
bands separated by two topologically non-trivial band gaps. We
observe the existence of one-way edge mode that spans these
topologically non-trivial band gaps, as shown in Figure
\ref{edge}b with $V = 0.5 \Omega$. Thus the one-way edge state
persist in the ultra-strong coupling regime.

For applications of one-way edge modes to carry information,
optimizing the bandwidth of such a one-wage edge mode is
important. Since the one-way edge mode spans the topologically
non-trivial band gap, the size of such a gap becomes a good
measure of the bandwidth of the one-way edge mode. In Figure
\ref{edge}a, we plot the size of the topologically non-trivial
band gap, as a function of modulation strength $V$. The bandwidth
increases with $V$ for small $V$, peaks at $V =0.5 \Omega$, and
then decreases to zero signifying the topological phase transition
mentioned above. For a given modulation frequency $\Omega$
therefore, the bandwidth of the one-way edge mode maximizes at the
ultra-strong coupling regime.

\begin{figure}[h]
\includegraphics[width=14cm]{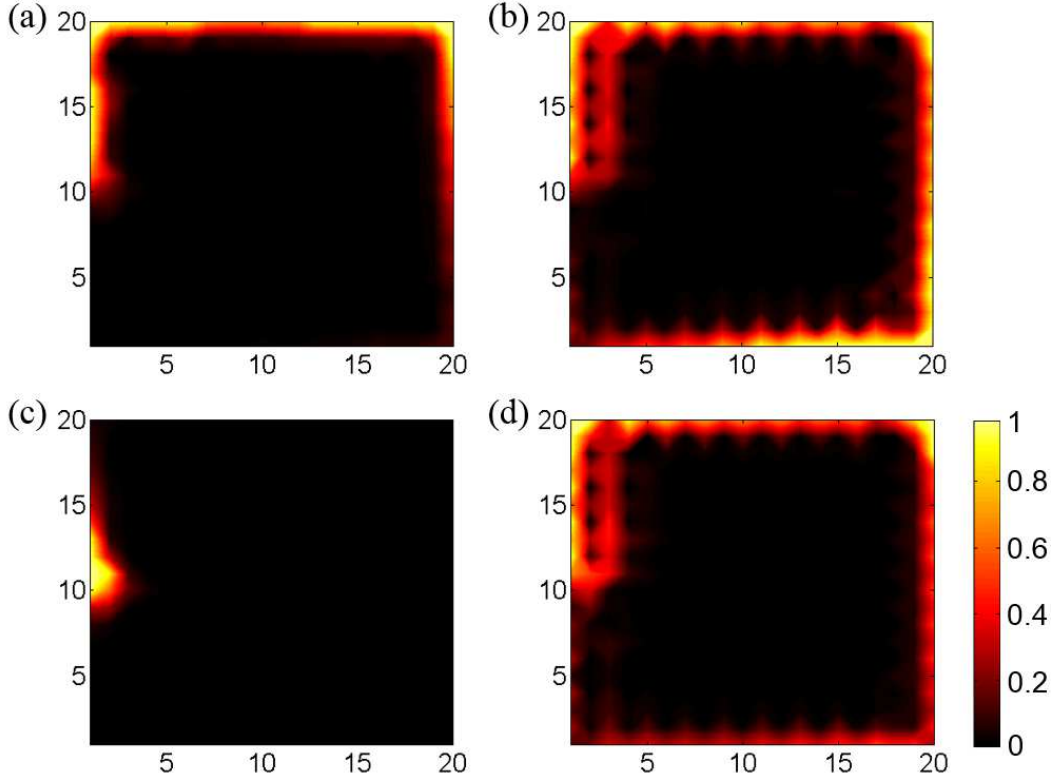}
\caption{The simulation results with the Hamiltonian (\ref{ham1})
in a $20\times20$ resonator lattice ($\omega_A = 12\pi c/a$,
$\omega_B = 0$). A point source is placed at the location (1,10).
The propagation field profiles are plotted at time (a) $t=100 a/c$
with $V=0.02\Omega$ and the source frequency $\omega_s=V$; (b)
$t=10 a/c$ with $V=0.5\Omega$ and $\omega_s=V$; (c) same as (a)
and (d) same as (b) with a loss coefficient $\gamma = 0.1 c/a$ at
steady state. \label{sim}}
\end{figure}

The one-way edge mode is topologically robust against
disorder-induced back-scattering. However, such a mode is still
susceptible to intrinsic losses of the materials. For practical
application then mitigation of the effect of intrinsic loss is
important. For a given modulation frequency $\Omega$, operating in
the ultra-strong coupling regime results in the one-way edge modes
that have larger group velocities, and hence are less susceptible
to loss, as compared to operating in the weak-coupling regime. As
an illustration, we consider a finite structure described by the
Hamiltonian of Eq. (\ref{ham1}) of a size of $20a$ by $20a$
(Figure \ref{sim}). We consider two systems in the weak-coupling
and ultra-strong coupling regime, corresponding to $V = 0.02
\Omega$, and $V = 0.5 \Omega$, respectively. To probe the systems,
we place a point source at the location ($1a$, $10a$), and choose
the frequencies of the point source to be inside one of the
topologically non-trivial band gaps. Both systems indeed support
one-way edge modes as shown in Figures \ref{sim}a and b. To study
the effect of loss, we further include into Eq. (\ref{ham1}) an
extra term $-\gamma \left( \sum_m a^\dagger_m a_m + \sum_n
b^\dagger_n b_n \right)$, where $\gamma = 0.1 c/a$ is the loss
coefficient. The steady-state field distributions for the same
source are shown in Figures \ref{sim}c and d respectively. We see
that the photons indeed have a much longer propagation distance in
the ultra-strong coupling regime with $V = 0.5 \Omega$ (Figure
\ref{sim}d), as compared to the weak-coupling regime with $V =
0.02 \Omega$ (Figure \ref{sim}c).

In summary, we consider a system of dynamically-modulated photonic
resonator lattice undergoing photonic transition, and show that
such a lattice can exhibit non-trivial topological properties in
the ultra-strong coupling regime. From an experimental and
practical point of view, for the same modulation frequency,
operating the system in the ultra-strong coupling regime results
in one-way edge modes that has a larger bandwidth, and is less
susceptible to loss, as compared to operating the same system in
the weak-coupling regime. Our work therefore should provide useful
guidance to the experimental quest in seeking to demonstrate
topological effects related to time-reversal symmetry breaking
on-chip \cite{tzuang14}. We also show that in the ultra-strong
coupling regime, the system undergoes a topological phase
transition as one varies the modulation strength. This phase
transition has no counter part in weak-coupling systems, and its
nature is directly related to the non-trivial topology of the
quasi-energy space. In the context of recent significant
fundamental interests in exploring the ultra-strong coupling
physics
\cite{irish05,irish07,rabl11,schiro12,burillo14,gunter09,imamoglu09,niemczyk10,scalari12,geiser12,li13,scalari14},
our work points to the exciting prospect of exploring non-trivial
topological effects in ultra-strong coupling regime.

\begin{acknowledgments}
This work is supported in part by U.S. Air Force Office of
Scientific Research Grant No. FA9550-12-1-0488 and U.S. National
Science Foundation Grant No. ECCS-1201914.
\end{acknowledgments}


\begin{thebibliography}{99}
\bibitem{klitzing80} K. V. Klitzing, G. Dorda, and M. Pepper,
\textit{Phys. Rev. Lett.} \textbf{45}, 494 (1980).
\bibitem{qi11} X.-L Qi, and S.-C. Zhang, \textit{Rev. Mod. Phys.} \textbf{83},
1057 (2011).
\bibitem{lu14} L. Lu, J. D. Joannopoulos, and M. Solja\v{c}i\'{c},
\textit{Nat. Photonics} \textbf{8}, 821 (2014).
\bibitem{haldane08} F. D. M. Haldane and S. Raghu, \textit{Phys. Rev.
Lett.} \textbf{100}, 013904 (2008).
\bibitem{raghu08} S. Raghu and F. D. M. Haldane, \textit{Phys. Rev.
A} \textbf{78}, 033834 (2008).
\bibitem{wang09} Z. Wang, Y. Chong, J. D. Joannopoulos, M.
Solja\v{c}i\'{c}, \textit{Nature} \textbf{461}, 772 (2009).
\bibitem{hafezi11} M. Hafezi, E. A. Demler, M. D. Lukin, and J. M.
Taylor, \textit{Nat. Phys.} \textbf{7}, 907 (2011).
\bibitem{poo11} Y. Poo, R. -X. Wu, Z. Lin, Y. Yang, and C. T.
Chan, \textit{Phys. Rev. Lett.} \textbf{106}, 093903 (2011).
\bibitem{hafezi13} M. Hafezi, S. Mittal, J. Fan, A. Migdall, and
J. M. Taylor, \textit{Nat. Photonics} \textbf{7}, 1001 (2013).
\bibitem{khanikaev13} A. B. Khanikaev, S. H. Mousavi, W. -K. Tse, M. Kargarian, A. H. MacDonald, and G. Shvets, \textit{Nature Materials} \textbf{12}, 233
(2013).
\bibitem{longhi13} S. Longhi, \textit{Opt. Lett.} \textbf{38},
3570 (2013).
\bibitem{liang13} G. Q. Liang and Y. D. Chong, \textit{Phys. Rev.
Lett.} \textbf{110}, 203904 (2013).
\bibitem{mittal14} S. Mittal, J. Fan, A. Faez, J. M. Taylor, and
M. Hafezi, \textit{Phys. Rev. Lett.} \textbf{113}, 087403 (2014).
\bibitem{li14} E. Li, B. J. Eggleton, K. Fang, and S. Fan,
\textit{Nat. Commun.} \textbf{5}, 3225 (2014).
\bibitem{slobozhanyuk15} A. P. Slobozhanyuk, A. N. Poddubny, A. E.
Miroshnichenko, P. A. Belov, and Y. S. Kivshar, \textit{Phys. Rev.
Lett.} \textbf{114}, 123901 (2015).
\bibitem{kitagawa10} T. Kitagawa, E. Berg, M. Rudner, and
E. Demler, \textit{Phys. Rev. B} \textbf{82}, 235114 (2010).
\bibitem{inoue10} J. Inoue and A. Tanaka, \textit{Phys. Rev.
Lett.} \textbf{105}, 017401 (2010).
\bibitem{lindner11} N. H. Lindner, G. Refael, and V. Galitski,
\textit{Nat. Phys.} \textbf{7}, 490 (2011).
\bibitem{leon13} A. G\'{o}mez-Le\'{o}n and G. Platero, \textit{Phys. Rev.
Lett.} \textbf{110}, 200403 (2013).
\bibitem{rudner13} M. S. Rudner, N. H. Lindner, E. Berg, and M. Levin,
\textit{Phys. Rev. X} \textbf{3}, 031005 (2013).
\bibitem{cayssol13} J. Cayssol, B. D\'{o}ra, F. Simon, and R.
Moessner, \textit{Phys. Status Solidi RRL} \textbf{7}, 101 (2013).
\bibitem{rechstman13} M. C. Rechstman, J. M. Zeuner, Y. Plotnik,
Y. Lumer, D. Podolsky, F. Dreisow, S. Nolte, M. Segev, and A.
Szameit, \textit{Nature} \textbf{496}, 196 (2013).
\bibitem{fang12np} K. Fang, Z. Yu, and S. Fan, \textit{Nat. Photonics} \textbf{6}, 782 (2012).
\bibitem{irish05} E. K. Irish, J. Gea-Banacloche, I. Martin, and
K. C. Schwab, \textit{Phys. Rev. B} \textbf{72}, 195410 (2005).
\bibitem{irish07} E. K. Irish, \textit{Phys. Rev. Lett.}
\textbf{99}, 173601 (2007).
\bibitem{rabl11} P. Rabl, \textit{Phys. Rev. Lett.} \textbf{107},
063601 (2011).
\bibitem{schiro12} M. Schir\'{o}, M. Bordyuh, B. \"{o}ztop, and H. E. T\"{u}reci, \textit{Phys. Rev. Lett.} \textbf{109},
053601 (2012).
\bibitem{burillo14} E. Sanchez-Burillo, D. Zueco, J. J. Garcia-Ripoll, and L. Martin-Moreno, \textit{Phys. Rev. Lett.} \textbf{113},
263604 (2014).
\bibitem{gunter09} G. G\"{u}nter, A. A. Anappara, J. Hees, A. Sell, G. Biasiol, L. Sorba, S. De Liberato, C. Ciuti, A. Tredicucci,
A. Leitenstorfer, and R. Huber \textit{Nature} \textbf{458}, 178
(2009).
\bibitem{imamoglu09} A. Imamo\u{g}lu, \textit{Phys. Rev. Lett.} \textbf{102},
083602 (2009).
\bibitem{niemczyk10} T. Niemczyk, F. Deppe, H. Huebl, E. P.
Menzel, F. Hocke, M. J. Schwarz, J. J. Garcia-Ripoll, D. Zueco, T.
H\"{u}mmer, E. Solano, A. Marx, and R. Gross, \textit{Nat. Phys.}
\textbf{6}, 772 (2010).
\bibitem{tzuang14} L. D. Tzuang, K. Fang, P. Nussenzveig, S. Fan, and M. Lipson, \textit{Nat. Photonics} \textbf{8}, 701 (2014).
\bibitem{winn99} J. N. Winn, S. Fan, J. D. Joannopoulos, and
E. P. Ippen, Phys. Rev. B \textbf{59}, 1551 (1999).
\bibitem{peierls33} R. E. Peierls, \textit{Z. Phys.} \textbf{80}, 763 (1933).
\bibitem{fang13} K. Fang and S. Fan, \textit{Phys. Rev. Lett.} \textbf{111}, 203901 (2013).
\bibitem{fang13oe} K. Fang, Z. Yu, and S. Fan, \textit{Opt. Express} \textbf{21}, 18216 (2013).
\bibitem{lin14} Q. Lin and S. Fan, \textit{Phys. Rev. X} \textbf{4}, 031031 (2014).
\bibitem{yuan15} L. Yuan and S. Fan, arXiv:1502.06037.
\bibitem{lipson05} M. Lipson, \textit{J. Lightwave Technol.}
\textbf{23} 4222 (2005).
\bibitem{scalari12} G. Scalari, C. Maissen, D.
Tur\v{c}inkov\'{a}, D. Hagenm\"{u}ller, S. De Liberato, C. Ciuti,
C. Reichl, D. Schuh, W. Wegscheider, M. Beck, J. Faist,
\textit{Science} \textbf{335}, 1323 (2012).
\bibitem{geiser12} M. Geiser, F. Castellano, G. Scalari, M. Beck,
L. Nevou, and J. Faist, \textit{Phys. Rev. Lett.} \textbf{108},
106402 (2012).
\bibitem{li13} J. Li, M. P. Silveri, K. S. Kumar, J. -M.
Pirkkalainen, A. Veps\"{a}l\"{a}inen, W. C. Chien, J. Tuorila, M.
A. Sillanp\"{a}\"{a}, P. J. Hakonen, E. V. Thuneberg, and G. S.
Paraoanu, \textit{Nat. Commun.} \textbf{4}, 1420 (2013).
\bibitem{scalari14} G. Scalari, C. Maissen, S. Cibella, R. Leoni,
P. Carelli, F. Valmorra, M. Beck, and J. Faist, \textit{New J.
Phys.} \textbf{16}, 033005 (2014).
\bibitem{shirley65} J. H. Shirley, \textit{Phys. Rev.}
\textbf{138}, B979 (1965).
\bibitem{samba73} H. Samba, \textit{Phys. Rev. A} \textbf{7}, 2203
(1973).
\end{thebibliography}
\end{document}